%% file: coxeter_np_algorithm.tex
\documentclass[draft, onecolumn]{IEEEtran}

\usepackage{mathbf-abbrevs}

\input{defs}

\begin{document}

\title{Linear-time nearest point algorithms for Coxeter lattices}

\author{Robby~G.~McKilliam,%
  \thanks{Robby~McKilliam is partly supported by a scholarship from
    the Wireless Technologies Laboratory, CSIRO ICT Centre, Sydney,
    Australia } Warren~D.~Smith 
    \thanks{Warren Smith is with the Center for Range Voting, 
    21 Shore Oaks Drive, Stony Brook NY 11790 USA} and I.~Vaughan~L.~Clarkson%
    \thanks{Robby~McKilliam and Vaughan~Clarkson are with the School of
    Information Technology \& Electrical Engineering, The University
    of Queensland, Qld., 4072, Australia}
}
\markboth{Robby~G.~McKilliam \emph{et al.}, Linear-time nearest point algorithms for Coxeter lattices}%
{DRAFT \today}

\maketitle

\begin{abstract}
The Coxeter lattices, which we denote $A_{n/m}$, are a family of lattices containing many of the important lattices in low dimensions.  This includes $A_n$, $E_7$, $E_8$ and their duals $A_n^*$, $E_7^*$ and $E_8^*$.  We consider the problem of finding a nearest point in a Coxeter lattice.  We describe two new algorithms, one with worst case arithmetic complexity $O(n\log{n})$ and the other with worst case complexity $O(n)$ where $n$ is the dimension of the lattice.  We show that for the particular lattices $A_n$ and $A_n^*$ the algorithms reduce to simple nearest point algorithms that already exist in the literature.
\end{abstract}


\section{Introduction}

The study of point lattices is of great importance in
several areas of number theory, particularly the studies of quadratic
forms, the geometry of numbers and simultaneous Diophantine
approximation, and also to the practical engineering problems of
quantisation and channel coding.  They are also important in studying
the sphere packing problem and the kissing number problem
\cite{Clarkson1999:Anstar, SPLAG}.  Lattices have recently found significant application of in cryptography \cite{AjtaiGeneratingHardInstancesOfLatticeProblems1996, AjtaiPublicKeyCryptoLattice1997} and communications systems using multiple antannaes \cite{Brunel_lattice_decoding_2003, Ryan2008}.

A \emph{lattice}, $L$, is a set of points in $\reals^n$ such that
\[ 
  L = \{\xbf \in \reals^n |  \xbf = \mathbf{Bw} , \wbf \in \mathbb Z^n \}
\]
where $\Bbf$ is termed the \emph{generator (or basis) matrix}.  We will write vectors and matrices in bold font.  The $i$th element in a vector is denoted by a subscript: $x_i$.  The generator matrix for a lattice is not unique.  Let $\Mbf$ be an $n \times n$ matrix with integer elements such that $\det(\Mbf) = \pm 1$.  $\Mbf$ is called a \emph{unimodular} matrix.  Then both $\Bbf$ and $\Bbf\Mbf$ are generator matrices for the lattice $L$.

Lattices are equivalent under scaling, rotation and reflection.  A lattice $L$ with generator matrix $\Bbf$ and a lattice $\hat{L}$ with generator lattice $\hat{\Bbf}$ are equivalent, or \emph{isomorphic}, iff
\[
\Bbf = \alpha \Rbf \hat{\Bbf} \Mbf
\]
where $\alpha > 0$ is real,   
$\Rbf$ is a matrix consisting of only rotations and reflections and $\Mbf$ is unimodular.  We write $L \simeq \hat{L}$.

The \emph{Voronoi region} or \emph{nearest-neighbour region} $\vor(L)$ for a lattice $L$ is the subset of $\reals^n$ such that, with respect to a given norm, all points in $\vor(L)$ are \emph{nearer} to the origin than to any other point in $L$.  
The Voronoi region is an $n$-dimensional 
polytope~\cite{SPLAG}.  Given some lattice point $\xbf \in L$ we will write $\vor(L) + \xbf$ to denote the Voronoi region centered around the lattice point $\xbf$.  It follows that $\vor(L) + \xbf$ is the subset of $\reals^n$ that is nearer to $\xbf$ than any other lattice point in $L$.   

The nearest lattice point problem is: Given $\ybf\in\reals^n$ and some lattice $L$ whose lattice points lie in~$\reals^n$, find a 
lattice point $\xbf \in L$ such that the Euclidean distance between $\ybf$ and $\xbf$ is minimised.  We use the notation $\NP(\ybf, L)$ to denote the nearest point to $\ybf$ in the lattice $L$.  It follows from the definition of the Voronoi region that\footnote{There is a slight technical deficiency here.  We actually require to define half of the faces of $\vor(L)$ to be closed and half to be open.  Ties in $\NP(\ybf, L)$ can then be broken accordingly.}
\[
	\xbf = \NP(\ybf, L) \Leftrightarrow \ybf \in \vor(L) + \xbf  
\]  
The nearest lattice point problem has significant practical application.  If the lattice is used for vector quantisation then the nearest lattice point corresponds to the minimum-distortion point.  If the lattice is used as a code for a Gaussian channel, then the nearest lattice point corresponds to maximum likelihood decoding \cite{Conway1982FastQuantDec}.  The closely related \emph{shortest vector problem} has been used in public key cryptography \cite{AjtaiGeneratingHardInstancesOfLatticeProblems1996, AjtaiPublicKeyCryptoLattice1997, feige_inapproximability_2004, regev_new_2004, Micciancio_lattice_based_post_quantum_crypto}.

Van Emde Boas \cite{vanEmdeBoas1981} and Ajtai \cite{Ajtai1998} have shown that the nearest lattice point problem is NP-complete under certain conditions when the lattice itself, or rather a basis thereof, is considered as an additional input parameter.  It has even been shown that finding approximately nearest points is NP-hard \cite{feige_inapproximability_2004, arora_hardness_1993, micciancio_worst-case_2004}.   Nevertheless, algorithms exist that can compute the nearest lattice point in reasonable time if the dimension is small \cite{Agrell2002, Viterbo_sphere_decoder_1999, Pohst_sphere_decoder_1981}. One such algorithm introduced by Pohst \cite{Pohst_sphere_decoder_1981} in 1981 was popularised in signal processing and communications fields by Viterbo and Boutros \cite{Viterbo_sphere_decoder_1999} and has since been called the \emph{sphere decoder}. 

For specific lattices, the nearest point problem is considerably easier and for many classical lattices, fast nearest point algorithms are known \cite{Conway1982FastQuantDec, Conway1986SoftDecLeechGolay, Clarkson1999:Anstar, McKilliam2008, McKilliam2008b, SPLAG}.

The Coxeter lattices, denoted $A_{n/m}$, are a family of lattices first described by H.S.M. Coxeter \cite{Coxeter1951, Martinet2003}.

\begin{equation} \label{eq:A_{n/m}}
A_{n/m} = \left\{ \Qbf\xbf \mid \xbf\in\ints^{n+1}, \, \xbf'\onebf \bmod {m} = 0 \right\}
\end{equation}
where $\Qbf$ is the orthogonal projection matrix
\begin{equation}\label{Q}
\Qbf =  \left(\Ibf - \frac{\bm{1} \bm{1}'}{n+1}\right),
\end{equation}
$\Ibf$ is the $(n+1)\times(n+1)$ identity matrix, $\onebf = [1, 1, 1, \dots]'$ and $'$ indicates the vector or matrix transpose.  If $m$ does not divide $n+1$ then $A_{n/m} = A_{n/1}$.  Hence, in the sequel, we assume that $m$ divides $n+1$.

A simple geometric description of $A_{n/m}$ is to consider the subset consisting of the points of $Z^{n+1}$ whose coordinate-sum is divisible by $m$.  This subset consists of points that lie in `layers' parallel to the hyperplane orthogonal to $\onebf$.  By projecting the subset orthogonally to $\onebf$ we obtain a set of points equivalent to the $n$-dimensinal lattice $A_{n/m}$. 

The family of Coxeter lattices contains many of the important lattices in low dimension.  The family is related to the well studied root lattice $A_n$ and its  
dual lattice $A_n^*$.  When $m = 1$
\begin{equation} \label{eq:A_{n/1}}
A_{n/1} = A_n^* = \{ \Qbf\xbf \mid \xbf \in \ints^{n+1} \}
\end{equation}
and when $m = n+1$
\begin{equation} \label{eq:A_{n/n+1}}
A_{n/n+1} = A_n = \left\{ \xbf\in\ints^{n+1} \mid \xbf'\onebf = 0 \right\}
\end{equation}
It follows that $A_n \subseteq A_{n/m} \subseteq A_n^*$~\cite{SPLAG, Martinet2003}.  Note that $A_{n/m} \subset A_{n/k}$ whenever $k < m$ and therefore
\begin{equation} \label{eq:VorAnksubsetAnm}
\vor(A_{n/k}) \subset \vor(A_{n/m}).
\end{equation} 


Other isomorphisms exist: $A_{8/3} \simeq E_8 \simeq E_8^*$, $A_{7/4} \simeq E_7$ and $A_{7/2} \simeq E_7^*$.  Of significant practical interest is the lattice $E_8 \simeq A_{8/3}$.  Due to its excellent packing and quatising properties $E_8$ has found applications to  trellis codes \cite{Calderbank1986E8TrellisCode, Wei1987TrellisCodedModulation, ForneyCosetCodes1, ForneyCosetCodes2} and vector quantisation \cite{SPLAG, Conway1982VoronoiRegions, Postol2002LatticeVideoCompression}.  The particular representation of $E_8$ as $A_{8/3}$ was used by Secord and deBuda to create a code with a spectral null at DC \cite{Secord1989}.

The lattice $A_n^* \simeq A_{n/1}$ is also of practical interest.  It gives the thinnest 
sphere-covering 
in all dimensions up to $8$~\cite{SPLAG} and has found application
in a number of estimation problems including period estimation from
sparse timing data \cite{Clarkson2007, McKilliam2007}, frequency estimation
\cite{Clarkson1999}, direction of arrival estimation
\cite{Quinn2007} and noncoherent detection \cite{McKilliam2009LinearTimeBlockPSK}.

The paper is organised as follows.  Section~\ref{sec:onlognalg} describes a log-linear-time nearest point algorithm for $A_{n/m}$.  This algorithm is a generalisation of a nearest point algorithm for $A_n^*$ that was derived in~\cite{McKilliam2008}.  Section~\ref{sec:onalg} improves this to 
worst case linear-time.
The speedup employs both a partial sorting procedure called a bucket sort~\cite{Cormen2001} and also the linear-time Rivest-Tarjan selection 
algorithm~\cite{Blum1973, Floyd1975, Rivest1975, KnuthACP2}. 
In Section~\ref{sec:otheralgorithmsAnAn*} we show how the discussed nearest point algorithms for the Coxeter lattices reduce to simple nearest point algorithms for $A_n$ and $A_n^*$ that already exist in the literature~\cite{SPLAG, McKilliam2008, McKilliam2008b}.  In Section \ref{sec:gluevectoralg} we review a simple nearest point algorithm for $A_{n/m}$ based on translates of the lattice $A_n$.  This algorithm was previously described by Conway and Sloane~\cite{Conway1982FastQuantDec, Conway1986SoftDecLeechGolay} but not directly applied to the Coxeter lattices.  The algorithm requires $O(n^2)$ arithmetic operations in the worst case.  In Section~\ref{sec:runtimean} we evaluate the practical computational performance of the algorithms.



\section{Log-linear-time algorithm} \label{sec:onlognalg}

In this section we describe a nearest point algorithm for $A_{n/m}$ that requires $O(n\log{n})$ operations in the worst case.  This algorithm is a generalisation of the nearest point algorithm for $A_n^*$ described in~\cite{McKilliam2008}.  To describe the algorithm we first require to derive some properties of the Voronoi region of $A_{n/m}$.  This is done in Lemmata \ref{lem:QVorZnsubsetVorAn} and \ref{lem:VorAn=QVorZn1}.  We firstly require the follow definitions.

Let $H$ be the hyperplane in $\reals^{n+1}$ orthogonal to $\onebf$.  $H$ is typically refered to as the \emph{zero-mean-plane}.  For some lattice $L$ we will use the notation $\vor_H(L)$ to denote the region $\vor(L) \cap H$.  For example $\vor_H(A_n)$ is the crossection of $\vor(A_n)$ lying in the hyperplane $H$.  Given some region $R \subset H$ we define the $n$-volume of $R$ as $\vol_H(R)$.  For example, the $n$-volume of $\vor_H(A_n)$ is denoted by $\vol_H(\vor_H(A_n))$.

Given a set of $n$-dimensional vectors $S$ and suitable matrix $\Mbf$ we will write $\Mbf S$ to denote the set with elements $\Mbf\sbf$ for all $\sbf\in S$.  For example $\Qbf\vor(\ints^{n+1})$ denotes the region of space that results from projecting $\vor(\ints^{n+1})$ onto the hyperplane $H$.

\begin{lemma} \label{lem:QVorZnsubsetVorAn}
\[
\Qbf\vor(\ints^{n+1}) \subseteq \vor_H(A_{n})
\]
\end{lemma}
\begin{IEEEproof}
Let $\ybf \in \vor(\ints^{n+1})$.  Decompose $\ybf$ into orthogonal components so that $\ybf = \Qbf \ybf + t \onebf$ for some $t \in \reals$.  Then $\Qbf\ybf \in \Qbf\vor(\ints^{n+1})$.  Assume that $\Qbf\ybf \notin \vor_H(A_n)$.  Then there exists some $\xbf \in A_n$ such that
\begin{align*}
\|\xbf - \Qbf\ybf\|^2 < \|\zerobf - \Qbf\ybf\|^2 & \Rightarrow \|\xbf - \ybf + t\onebf\|^2 < \|\ybf - t\onebf\|^2 \\
& \Rightarrow \|\xbf - \ybf\|^2 + 2t\xbf'\onebf < \|\ybf\|^2.
\end{align*}
By definition \eqref{eq:A_{n/n+1}} $\xbf'\onebf = 0$ and so $\|\xbf - \ybf\|^2 < \|\ybf\|^2$.  This violates that $\ybf \in \vor(\ints^{n+1})$ and hence $\Qbf\ybf \in \vor_H(A_n)$.
\end{IEEEproof}

\begin{lemma} \label{lem:VorAn=QVorZn1}
\[
\vor_H(A_{n/m}) \subseteq \Qbf\vor(\ints^{n+1})
\]
with equality only when $m = n+1$.
\end{lemma}
\begin{IEEEproof}
When $m = n+1$, $A_{n/n+1} = A_n$.  The $n$-volume $\vol_H(\vor_H(A_n)) = \sqrt{n + 1}$ \cite{SPLAG}.  From Berger \emph{et al.} \cite{Burger1996} we find that the $n$-volume of the projected polytope $\vol_H(\Qbf\vor(\ints^{n+1})) = \sqrt{n + 1}$ also.  As $\vor_H(A_n)$ and $\Qbf\vor(\ints^{n+1})$ are convex polytopes it follows from \Lem{lem:QVorZnsubsetVorAn} that 
\[
\vor_H(A_n) = \Qbf\vor(\ints^{n+1}).
\]
The proof follows from the fact that $\vor_H(A_{n/m}) \subseteq \vor_H(A_n)$ for all $m$ \eqref{eq:VorAnksubsetAnm}.
\end{IEEEproof}

We will now prove Lemma \ref{lem:closest} from which our algorithm is derived.  We firstly need the following definition.  Given two sets $A$ and $B$ we let $A+B$ be their Minkowski sum.  That is, $x \in A + B$ iff $x = a + b$ where $a \in A$ and $b \in B$.  We will also write $\onebf\reals$ to denote the line of points $\onebf r$ for all $r \in \reals$.  Then $\vor_H(A_{n/m}) + \onebf\reals$ is an infinite cylinder with cross-section $\vor_H(A_{n/m})$.  It follows that $\vor_H(A_{n/m}) + \onebf\reals = \vor(A_{n/m})$

\begin{lemma} \label{lem:closest}
  If $\xbf = \Qbf \kbf$ is a closest point in $A_{n/m}$ to $\ybf \in
  \reals^{n+1}$ then there exists some $\lambda \in \reals$ for which
  $\kbf$ is a closest point in $\integers^{n+1}$ to $\ybf + \lambda
  \onebf$.
\end{lemma}

\begin{IEEEproof}
As $\Qbf \kbf$ is the nearest point to $\ybf$ then for all $\lambda\in\reals$
\[
\ybf + \onebf\lambda \in \vor(A_{n/m}) + \Qbf\kbf = \vor_H(A_{n/m}) + \kbf + \onebf\reals.
\]  
It follows from \Lem{lem:VorAn=QVorZn1} that
\[
\vor_H(A_{n/m}) + \kbf + \onebf\reals \subseteq \Qbf\vor(\ints^{n+1}) + \kbf + \onebf\reals.
\]
Then $\ybf + \onebf\lambda \in \Qbf\vor(\ints^{n+1}) + \kbf + \onebf\reals$ and for some $\lambda\in\reals$
\[
\ybf + \onebf\lambda \in \vor(\ints^{n+1}) + \kbf
\]
The proof now follows from the definition of the Voronoi region.
\end{IEEEproof}

Now consider the function $\fbf : \reals \mapsto \integers^{n+1}$
defined so that
\begin{equation} \label{eq:f}
  \fbf(\lambda) = \lfloor \ybf + \lambda \onebf \rceil
\end{equation}
where $\lfloor \cdot \rceil$ applied to a vector denotes the vector in which each element is rounded to a nearest integer\footnote{The direction of rounding for half-integers is not important so long as it's consistent.  The authors have chosen to round up half-integers in their own implementation.}.  That is, $\fbf(\lambda)$ gives a
nearest point in $\integers^{n+1}$ to $\ybf + \lambda \onebf$ as a function of $\lambda$.  Observe that $\fbf(\lambda + 1) = \fbf(\lambda) + \onebf$.  Hence,
\begin{equation}
  \Qbf \fbf(\lambda + 1) = \Qbf \fbf(\lambda).  \label{Qfl+1=Qfl}
\end{equation}

Lemma~\ref{lem:closest} implies there exists some $\lambda \in \reals$
such that $\xbf = \Qbf \fbf(\lambda)$ is a closest point to $\ybf$.
Furthermore, we see from~(\ref{Qfl+1=Qfl}) that $\lambda$ can be found
within an interval of length 1.  Hence, if we define the set
\begin{equation*}
  S = \{ \fbf(\lambda) \mid \lambda \in [0, 1)\}
\end{equation*}
then $\Qbf S$ contains a closest point in $A_{n/m}$ to $\ybf$.  In order to evaluate the elements in $S$ we require the following function.
\begin{definition} \label{def:sortindicies} \textbf{\emph{(sort indices)}}

We define the function
\[
\sbf = \sortindicies(\zbf)
\]
to take a vector $\zbf$ of length $n+1$ and return a vector $\sbf$ of indices such that
\[
z_{s_1} \geq z_{s_2} \geq z_{s_3} \geq \dots \geq z_{s_{n+1}}
\]
\end{definition}

Let
\[
\sbf = \sortindicies(\fracpart{\ybf})
\]
where $\fracpart{g} = g - \round{g}$ denotes the centered fractional part of $g\in\reals$ and we define $\fracpart{\cdot}$ to operate on vectors by taking the centered fractional part of each element in the vector.
It is clear that $S$ contains at most $n+2$ vectors, \emph{i.e.},
\begin{align} \label{scrS}
  S \subseteq \big\{\round{\ybf},
        \round{\ybf} + \ebf_{s_1}, \round{\ybf} + \ebf_{s_1} + \ebf_{s_2},
        \dots, \;\;\;\;\;\; \nonumber \\ 
        \round{\ybf} + \ebf_{s_1} + \dots + \ebf_{s_{n+1}}
 \big\}
\end{align}
where $\ebf_{i}$ is a vector of 0's with a 1 in the $i$th position. It can be seen that the last vector listed in the set is simply
$\lfloor \ybf \rceil + \onebf$ and so, once multiplied by $\Qbf$, the
first and the last vector are identical.

We can define the set $W \subseteq  S$ such that
\begin{equation} \label{eq:scrW}
W = \{ \xbf \in S \mid \xbf\cdot\onebf \bmod {m} = 0 \}.
\end{equation}
Noting \eqref{eq:A_{n/m}} then $\Qbf W$ contains the nearest point in $A_{n/m}$ to $\ybf$.

An algorithm suggests itself: test each of the distinct vectors in $\Qbf W$ and find the closest one to $\ybf$.  This is the principle of the algorithm we propose in this Section.  It remains to show that this can be done in $O(n \log n)$ arithmetic operations.

We label the elements of $S$ according to the order given
in~\eqref{scrS}.  That is, we set $\ubf_0 = \round{\ybf}$ and, for $i
= 1, \dots, n$,
\begin{equation}
  \ubf_i = \ubf_{i-1} + \ebf_{s_i} \label{eq_updateS}.
\end{equation}
Let $\zbf_i = \ybf - \ubf_i$.  Clearly, $\zbf_0 = \{\ybf\}$. Decompose $\ybf$ into orthogonal components $\Qbf \ybf$ and $t\onebf$ for some $t \in \reals$. The squared distance between $\Qbf \ubf_i$
and $\ybf$ is
\begin{equation}
\|\ybf - \Qbf\ubf_i\|^2 = d_i + t^2(n+1) \label{fastd}
\end{equation}
where we define $d_i$ as
\begin{equation}
  d_i = \|\Qbf \zbf_i\|^2
        = \left\|\zbf_i - \frac{\zbf_{i}' \onebf}{n + 1} \onebf \right\|^2
        = \zbf_{i}'\zbf_{i} - \frac{(\zbf_{i}' \onebf)^2}{n+1}.
\end{equation}
We know that the nearest point to $\ybf$ is that $\Qbf \ubf_i$ such that $\ubf_i \in W$ which minimizes~\eqref{fastd}.  Since the term $t^2(n+1)$ is independent of the index $i$, we can ignore it.  That is, it is sufficient to
minimize $d_i$, $i = 0, \dots, n$.

We now show that $d_i$ can be calculated inexpensively in a recursive
fashion.  We define two new quantities, $\alpha_i = \zbf_{i}' \onebf$
and $\beta_i = \zbf_{i}' \zbf_i$.  Clearly $d_i = \beta_i - \nicefrac{\alpha_i^2}{n+1}$.  From~\eqref{eq_updateS},
\begin{equation}
  \alpha_i = \zbf_{i}' \onebf = (\zbf_{i-1} - \ebf_{s_i})' \onebf = \alpha_{i-1} - 1 \label{update_zt1}
\end{equation}
and
\begin{equation}
  \beta_i = \zbf_{i}' \zbf_i = (\zbf_{i-1} - \ebf_{s_i})' (\zbf_{i-1} - \ebf_{s_i}) = \beta_{i-1} - 2\{y_{s_i}\} + 1. \label{update_ztz}
\end{equation}

Algorithm~\ref{alg:loglinearAnm} now follows.  The
main loop beginning at line~\ref{alg_for_all_bres} calculates the
$\alpha_i$ and $\beta_i$ recursively.  There is no need to retain
their previous values, so the subscripts are dropped.  The variable
$D$ maintains the minimum value of the (implicitly calculated values
of) $d_i$ so far encountered, and $k$ the corresponding index.  The variable $\gamma$ maintains the value of $\ubf_i'\onebf \bmod m$ which must equal $0$ in order for $\ubf_i \in W$.

Each line of the main loop requires $O(1)$ arithmetic computations so the loop (and that on line~\ref{alg_for_2}) requires $O(n)$ in total. The function $\sortindicies(\zbf)$ requires sorting $n+1$ elements.  This requires $O(n\log{n})$ arithmetic operations.  The vector operations on lines~\ref{alg_z}--\ref{alg_gamma} all require $O(n)$ operations and the matrix multiplication on line~\ref{alg_project} can be performed in $O(n)$ operations as
\[
\Qbf\ubf = \ubf - \frac{\onebf'\ubf}{n+1} \onebf.
\]
It can be seen, then, that the computational cost of the algorithm is dominated by the $\sortindicies(\cdot)$ function and is therefore $O(n \log n)$.

This algorithm is similar to the nearest point algorithm for $A_n^*$ described in \cite{McKilliam2008}.  The significant difference is the addition of $\gamma = 0$ on line $\ref{alg_if}$.  This ensures that the lattice points considered are elements of $A_{n/m}$ i.e. they satisfy \eqref{eq:A_{n/m}}.  We further discuss the relationship between the algorithms in Section~\ref{sec:otheralgorithmsAnAn*}.

\begin{algorithm} \label{alg:loglinearAnm}
\SetAlCapFnt{\small}
\SetAlTitleFnt{}
\caption{Algorithm to find a nearest lattice point in $A_{n/m}$ to
  $\ybf\in\reals^{n+1}$ that requires $O(n\log{n})$ arithmetic operations}
\dontprintsemicolon
\KwIn{$\ybf \in \reals^{n+1}$}
$\ubf = \round{\ybf}$ \nllabel{alg_k} \;
$\zbf = \ybf - \ubf$ \nllabel{alg_z}\;
$\alpha = \zbf' \onebf$ \nllabel{alg_alpha}\;
$\beta = \zbf'\zbf$ \nllabel{alg_beta}\;
$\gamma = \ubf'\onebf \bmod m$ \nllabel{alg_gamma} \;
$\sbf = \sortindicies(\zbf)$ \nllabel{alg_sortindices}\;
$D = \infty$ \;
\For{$i = 1$ \emph{\textbf{to}} $n+1$ \nllabel{alg_for_all_bres}}{
 \If{$\beta - \frac{\alpha^2}{n+1} < D$ \emph{\textbf{and}} $\gamma = 0$ \nllabel{alg_if}}{
 	$D = \beta - \frac{\alpha^2}{n+1}$ \;
 	$k = i-1$ \;
 }
 $\alpha = \alpha - 1$ \nllabel{alg_upalpha} \;
 $\beta = \beta - 2 z_{s_i} + 1$  \nllabel{alg_upbeta} \;
 $\gamma = (\gamma + 1) \bmod m$ \;
}
\For{$i = 1$ \emph{\textbf{to}} $k$ \nllabel{alg_for_2}}{
$u_{s_i} = u_{s_i} + 1$ \;
}
$\xbf = \Qbf\ubf$ \nllabel{alg_project}\;
\Return{$\xbf$ \nllabel{alg_return}}
\end{algorithm}

\section{Linear-time algorithm} \label{sec:onalg}

In the previous Section we showed that the nearest point to $\ybf$ in $A_{n/m}$ lies in the set $\Qbf W$ \eqref{eq:scrW}.  We will show that some of the elements of $\Qbf W$ can be immediately excluded from consideration.  This property leads to a nearest point algorithm that requires at most $O(n)$ arithmetic operations.

\begin{lemma} \label{lem:only2perbucket}
Suppose, for some integers $i, m > 0, k \geq 2$, that
\begin{equation} \label{eq:only2perbucket_condition}
\{y_{s_i}\} - \{y_{s_{i + km}}\} \leq \frac{m}{n+1}.
\end{equation}
Then the minimum of the $d_{i + cm}$, $c = 0, \dots, k$, occurs at $c = 0$ or $c = k$.
\end{lemma}
\begin{IEEEproof}
The proof proceeds by contradiction.  Suppose, to the contrary, that
\begin{equation*}
  d_{i+cm} < d_i \qquad \text{and} \qquad d_{i+cm} < d_{i+km}.
\end{equation*}
Observe that
\begin{equation*}
  d_{i+cm} - d_i = \frac{2 \alpha_i c m - \br{cm}^2}{n+1} + \sum_{j=1}^{cm} (1 - 2 \{y_{s_{i+j}}\}).
\end{equation*}
Now, since $\cubr{y_{s_{i+j}}} \leq \cubr{y_{s_i}}$, it follows that
\begin{equation*}
  d_{i+cm} - d_i \geq \frac{2 \alpha_i c m - \br{cm}^2}{n+1} + cm (1 - 2 \{y_{s_i}\})
\end{equation*}
With the assumption that $d_{i+cm} - d_i < 0$, we have that
\begin{equation}
  \frac{2 \alpha_i - cm}{n+1} < 2 \{y_{s_i}\} - 1. \label{gt}
\end{equation}
Similarly, observe that
\begin{align*}
  d_{i+km} - d_{i+cm} = \frac{2 \alpha_i (k-c) m - (k^2 - c^2) m^2}{n+1} \\
    + \sum_{j = cm+1}^{km} (1 - 2 \{y_{s_{i+j}}\}).
\end{align*}
Since $\cubr{y_{s_{i+j}}} \geq \cubr{y_{s_{i+km}}}$, it follows that
\begin{align*}
  d_{i+km} - d_{i+cm} \leq \frac{2 \alpha_i \br{k-c} m - \br{k^2 - c^2} m^2}{n+1} \\
    + \br{k-c} m \br{1 - 2 \cubr{y_{s_{i+km}}}}.
\end{align*}
With the assumption that $d_{i+km} - d_{i+cm} > 0$, we have that
\begin{equation}
  \frac{2 \alpha_i - cm}{n+1}> \frac{km}{n+1} - 1 + 2 \cubr{y_{s_{i+km}}}.  \label{lt}
\end{equation}
Equations~\refeqn{gt} and~\refeqn{lt} together imply that
\begin{equation*}
  \cubr{y_{s_i}} - \cubr{y_{s_{i+km}}} > \frac{km}{2 \br{n+1}},
\end{equation*}
which contradicts~\refeqn{eq:only2perbucket_condition} because $k \geq 2$.
\end{IEEEproof}

From $S$ we can construct the following $q = \nicefrac{n+1}{m}$ subsets
\begin{equation} \label{eq:U_j}
U_j = \left\{ \ubf_i \mid 0.5 - \{y_{s_i}\} \in \left(\frac{m(j-1)}{n+1}, \frac{mj}{n+1} \right] \right\}
\end{equation}
where $j=1,\cdots,q$.  Note that $\Qbf S = \Qbf\bigcup_{j=1}^{q}{U_j}$.  We are interested in the elements of $U_j\cap W$.  Let $g$ be the smallest integer such that $\ubf_g \in U_j \cap W$.  Let $p$ be the largest integer such that $\ubf_p \in U_j \cap W$.  It follows that $p = g + km$ for some $k \in \ints$.  Also, from \eqref{eq:U_j}
\[
\{y_{s_g}\} - \{y_{s_p}\} \leq \frac{m}{n+1}.
\]
It then follows from \Lem{lem:only2perbucket} that \eqref{fastd} is minimised either by $\ubf_g$ or $\ubf_p$ and not by any $\ubf_i \in U_j \cap W$ where $g < i < p$.  We see that for each set $\Qbf U_j$ there are at most two elements that are candidates for the nearest point.  An algorithm can be constructed as follows:  test the (at most two) candidates in each set $\Qbf U_j$ and return the closest one to $\ybf$.  We will now show how this can be achieved in linear time. 

%

We construct $q$ sets 
\begin{equation} \label{eq:B_j}
 B_j = \left\{ i \mid 0.5 - \{y_i\} \in \left(\frac{m(j-1)}{n+1}, \frac{mj}{n+1} \right] \right\}.
\end{equation}
and the related sets
\[
K_j = \bigcup_{t=1}^{j} B_t.
\]
It follows that
\[
\ubf_{|K_j|} = \round{\ybf} + \sum_{t \in K_j}{ \ebf_t }.
\]

\begin{definition} \label{def:quickselect} \textbf{\emph{(quick partition)}}

We define the function
\[
\bbf = \quickpartition(\zbf, B_j, c)
\]
to take a vector $\zbf$ and integer $c=1,\dots,|B_j|$ and return a vector $\bbf$ of length $|B_j|$ such that for $i = 1, \dots, c-1$ and $t = c+1, \dots, |B_j|$
\[
z_{b_i} \geq z_{b_c} \geq z_{b_t}
\]
\end{definition}
Somewhat surprisingly $\quickpartition(\zbf, B_j, c)$ can be implemented such that the required number of operations is $O(|B_j|)$.  This is facilitated by the Rivest-Tarjan selection algorithm~\cite{Blum1973, Floyd1975, Rivest1975, KnuthACP2}.  We can compute
\begin{equation} \label{eq:A=largest()}
\bbf = \quickpartition(\zbf, B_j, c)
\end{equation}
for some integer $1 \leq c \leq |B_j|$.  Then 
\begin{equation} \label{eq:u_|K|+c=round(y)+sum}
\ubf_{|K_{j-1}|+c} = \ubf_{|K_{j-1}|} + \sum_{t \in 1}^{c}{ \ebf_{b_t} }.
\end{equation}
Let $g$ be the smallest integer such that $1 \leq g \leq |B_j|$ and
\begin{equation} \label{eq:gmod0}
\onebf \cdot \ubf_{|K_{j-1}| + g} \bmod m = 0
\end{equation}
and let $p$ be the largest integer such that $1 \leq p \leq |B_j|$ and
\begin{equation} \label{eq:pmod0}
\onebf \cdot \ubf_{|K_{j-1}| + p} \bmod m = 0.
\end{equation}
From the previous discussion the only candidates for the nearest point out of the elements
\[
\Qbf\left\{\ubf_{|K_{j-1}|+1}, \dots, \ubf_{|K_{j-1}|+|B_j|} \right\} = \Qbf U_j
\]
are $\Qbf\ubf_{|K_{j-1}| + g}$ and $\Qbf\ubf_{|K_{j-1}| + p}$.  We can compute these quickly using the $\quickpartition(\cdot)$ function as in \eqref{eq:A=largest()} and \eqref{eq:u_|K|+c=round(y)+sum}.

\Alg{alg:slow} now follows.  Lines \ref{alg:slow:empty_buckets}-\ref{alg:slow:create_buckets_last_line} construct the sets $B_j$.  The main loop on line \ref{alg:slow:for_all_bres} then computes the values of $g$ and $p$ for each $B_j$.  We define the function 
\[
\bbf = \quickpartitiontwo(\zbf, B_j, g, p)
\]
to return $\bbf$ so that for $i = 1, \dots, g-1$ and $t = g+1, \dots, p-1$ and $c = p+1, \dots, |B_j|$
\[
z_{b_i} \geq z_{b_g} \geq z_{b_t} \geq z_{b_p} \geq z_{b_c}.
\]
Notice that $\quickpartitiontwo(\cdot)$ can be performed by two consecutive iterations of the Rivest-Tarjan algorithm and therefore requires $O(|B_j|)$ operations.  The $d_{|K_j| + g}$ and $d_{|K_j| + p}$ are computed within the loop on line \ref{alg:slow:inner_loop_test} and the index of the nearest lattice point is stored using the variable $k^*$.  The $\operatorname{concatenate}(\wbf, \bbf)$ function on line \ref{alg:slow:concatenate} adds the elements of $\bbf$ to the end of the array $\wbf$.  This can be performed in $O(|B_j|)$ operations.  Lines \ref{alg:slow:recoverpoint}--\ref{alg:slow:project} recovers the nearest lattice point using $\wbf$ and $k^*$.

In practice the $B_j$ can be implemented as a list so that the set insertion operation on line \ref{alg:slow:create_buckets_last_line} can be performed in constant time.  Then the loops on lines \ref{alg:slow:empty_buckets} and~\ref{alg:slow:create_buckets} require $O(n)$ arithmetic operations.  The operations inside the main loop on line~\ref{alg:slow:for_all_bres} require $O(|B_j|)$ operations.  The complexity of these loops is then
\[
\sum_{j=1}^{\nicefrac{n+1}{m}}{O(|B_j|)} = O(n)
\]
The remaining lines require $O(n)$ or less operations.  The algorithm then requires $O(n)$ arithmetic operations.

\begin{algorithm} \label{alg:slow}
\SetAlCapFnt{\small}
\SetAlTitleFnt{}
\caption{Algorithm to find a nearest lattice point in $A_{n/m}$ to
  $\ybf\in\reals^{n+1}$ that requires $O(n)$ arithmetic operations}
\dontprintsemicolon
\KwIn{$\ybf \in \reals^{n+1}$}
$\zbf = \ybf - \round{\ybf}$ \nllabel{alg:slow:z}\;
\lFor{$j = 1$ \emph{\textbf{to}} $q$ \nllabel{alg:slow:empty_buckets} }{$B_j = \emptyset$ \;}
\For{$i = 1$ \emph{\textbf{to}} $n+1$ \nllabel{alg:slow:create_buckets}}{
$j = q - \floor{q(z_i + \nicefrac{1}{2})}$ \;
$B_j = B_j \cup i$ \nllabel{alg:slow:create_buckets_last_line} \; 
}
$\ubf = \round{\ybf}$ \;
$\alpha = \zbf' \onebf$ \nllabel{alg:slow:alpha}\;
$\beta = \zbf'\zbf$ \nllabel{alg:slow:beta}\;
$\gamma = \ubf'\onebf \bmod m$ \nllabel{alg:slow:gamma} \;
$k = 1$ \;
$D = \infty$ \;
\For{$j = 1$ \emph{\textbf{to}} $q$ \nllabel{alg:slow:for_all_bres}}{

	$g = m - \gamma$ \;	
	$p = |B_j| - (|B_j| + \gamma) \bmod m$ \;
	$\bbf = \quickpartitiontwo(\zbf, B_j, g, p)$ \;
 	
	\For{$i = 1$ \emph{\textbf{to}} $|B_j|$ \nllabel{alg:slow:inner_loop_test}}{
		$\alpha = \alpha - 1$ \;
		$\beta = \beta - 2 z_{b_i} + 1$ \;
		$\gamma = \br{\gamma + 1} \bmod m$ \;
		\If{($i = g$ \emph{\textbf{or}} $i = p$) \emph{\textbf{and}} $\beta - \nicefrac{\alpha^2}{n+1} < D$}{
			$D = \beta - \alpha^2 / \br{n+1}$ \;
			$k^* = k$ \;
		}
		$k = k+1$ \;
	}
	$\operatorname{concatenate}(\wbf, \bbf)$ \nllabel{alg:slow:concatenate} \;
}
\For{$i = 1$ \emph{\textbf{to}} $k^*$ \nllabel{alg:slow:recoverpoint}}{
    $u_{w_i} = u_{w_i} + 1$ \;
}
  $\xbf = \Qbf \ubf$ \nllabel{alg:slow:project} \;
\Return{$\xbf$ \nllabel{alg:slow:return}}
\end{algorithm}

\section{Specific algorithms for $A_n$ and $A_n^*$} \label{sec:otheralgorithmsAnAn*}

For the lattices $A_n = A_{n/n+1}$ and $A_n^* = A_{n/1}$ Algorithms \ref{alg:loglinearAnm} and \ref{alg:slow} reduce to simpler algorithms that have previously been described in the literature.  For $A_n$ a log-linear time algorithm similar to that of Conway and Sloane~\cite{Conway1982FastQuantDec,Conway1986SoftDecLeechGolay} is derived from Algorithm \ref{alg:loglinearAnm} by noting that only one iteration in the main loop on line \ref{alg_for_all_bres} will satisfy $\gamma = 0$.  Algorithm \ref{alg:Anonlogn} now follows.

\begin{algorithm} \label{alg:Anonlogn}
\SetAlCapFnt{\small}
\SetAlTitleFnt{}
\KwIn{$\ybf \in \reals^{n+1}$}
\dontprintsemicolon
\caption{Algorithm to find a nearest lattice point in $A_n$ to $y \in \reals^n$ that requires $O(n\log{n})$ operations}
$\gamma = (n+1 - \round{\ybf}'\onebf) \; \text{mod} \; n+1$ \;
$\sbf = \sortindicies(\fracpart{\ybf})$ \nllabel{alg:Anonlogn:sortindicies} \;
$\ubf = \round{\ybf}$ \;
\ForEach{ $i = 1$ \emph{\textbf{to}} $\gamma$ }{
$u_{s_i} = u_{s_i} + 1$ \;
}
$\xbf = \Qbf\ubf$ \nllabel{alg:Anonlogn:project}\;
\Return{$\xbf$}
\end{algorithm}

A simple linear-time algorithm for $A_n$ can be constructed from \Alg{alg:Anonlogn} by replacing the $\sortindicies(\cdot)$ function on line~\ref{alg:Anonlogn:sortindicies} with $\quickpartition(\cdot)$.  Pseudocode is provided in \Alg{alg:Anon}.  In effect this is a modification of \Alg{alg:slow} where the sets from \eqref{eq:B_j} are replaced by the single set $\{1,2,\dots,n+1\}$.  This algorithm has previously been suggested by A. M. Odlyzko \cite[page 448]{SPLAG}.

\begin{algorithm} \label{alg:Anon}
\SetAlCapFnt{\small}
\SetAlTitleFnt{}
\KwIn{$\ybf \in \reals^{n+1}$}
\dontprintsemicolon
\caption{Algorithm to find a nearest lattice point in $A_n$ to $y \in \reals^n$ that requires $O(n)$ operations}
$\gamma = (n+1 - \round{\ybf}'\onebf) \; \text{mod} \; n+1$ \;
$\bbf = \quickpartition(\fracpart{\ybf}, \{1,2,\dots,n+1\}, \gamma)$ \;
$\ubf = \round{\ybf}$ \;
\For{ $i = 1$ \emph{\textbf{to}} $\gamma$ }{
$u_{b_i} = u_{b_i} + 1$ \;
}
$\xbf = \Qbf\ubf$ \nllabel{alg:An:project}\;
\Return{$\xbf$}
\end{algorithm}

For $A_n^*$ a log-linear time algorithm identical to that described in \cite{McKilliam2008} can be derived from Algorithm \ref{alg:loglinearAnm} by noting that $\gamma \bmod {1} = 0$ for all $\gamma$.  A linear-time algorithm for $A_n^*$ can be constructed from \Alg{alg:slow} by noting that $g = 1$ \eqref{eq:gmod0} and $p = |B_j|$ \eqref{eq:pmod0} for all $B_j$ where $j = 1, 2, \dots, n+1$.  This removes the need for using the $\quickpartitiontwo(\cdot)$ function.  A further simplification is noted in~\cite{McKilliam2008b} where it was shown that the nearest point is one of the $\Qbf\ubf_{|K_j|}$ where $j=0,\cdots,n$.  The reader is referred to \cite{McKilliam2008b} for further details.  The proofs used in \cite{McKilliam2008b} are significantly different to those in this paper and are only applicable to $A_n^*$.


\section{Algorithm based on glue vectors} \label{sec:gluevectoralg}

In this section we describe a simple nearest point algorithm for $A_{n/m}$.  This algorithm was described by Conway and Sloane~\cite{Conway1982FastQuantDec, Conway1986SoftDecLeechGolay} but not directly applied to the Coxeter lattices.  The algorithm has worst case complexity $O(n^2)$.

$A_{n/m}$ can be constructed by \emph{gluing} translates of the lattice $A_{n}$~\cite{SPLAG}.  That is
\begin{equation}
  A_{n/m} = \bigcup_{i=0}^{q-1} \left( [im] + A_{n} \right) \label{eq:[i]Anm}
\end{equation}
where $q = \nicefrac{n+1}{m}$ and $[i]$ are called \emph{glue vectors} and are defined as
\begin{equation} \label{eq:Anglues}
  [i] = \frac{1}{n+1} \big(\underbrace{i, \dots, i}_{\text{$j$ times}},
        \underbrace{-j, \dots, -j}_{\text{$i$ times}}
  \big)
\end{equation}
for $i \in \{0,\dots, n\}$ with $i+j = n+1$.  Following the notation of Conway and Sloane the glue vectors will not be written in boldface.  Instead they are indicated by square brackets.

Noting that $A_{n/m}$ can be constructed as a union of $q$ translates of the lattice $A_n$ we can use a nearest point algorithm for $A_n$ to find the nearest point in each of the translates.  The translate containing the closest point yields the nearest point in $A_{n/m}$.  A pseudocode implementation is provided in \Alg{alg:gluealgorithm}.  The function $\NP(\ybf, A_n)$ can be implemented by either Algorithm \ref{alg:Anonlogn} or \ref{alg:Anon} of Section \ref{sec:otheralgorithmsAnAn*}.

\begin{algorithm} \label{alg:gluealgorithm}
\SetAlCapFnt{\small}
\SetAlTitleFnt{}
\dontprintsemicolon
\KwIn{$\ybf \in \reals^n$}
$D = \infty$ \;
\For{$i = 0$ \emph{\textbf{to}} $q - 1$}{ 
	    $\xbf = \NP(\ybf - [im], A_n) + [im]$ \;
	    \If{$\|\xbf - \ybf\| < D$}
	    { 
	    	$\xbf_{\text{NP}} = \xbf$ \;
	    	$D = \|\xbf - \ybf\|$ \;
	    }
}
\Return{ $\xbf_{\text{\emph{NP}}}$ }
\caption{Nearest point algorithm for $A_{n/m}$ using glue vectors}
\label{alg_np_gluevectors}
\end{algorithm}

The algorithm requires iterating $\NP(\ybf, A_n)$ $q$ times.  Assuming that $\NP(\ybf, A_n)$ is implemented using the linear time algorithm (Algorithm \ref{alg:Anon}) then if $q$ is a constant this yields a linear-time algorithm.  At worst $q$ may grow linearly with $n$.  In this case the algorithm requires $O(n^2)$ operations.

\section{Run-time analysis} \label{sec:runtimean}

In this section we tabulate some practical computation times attained with the nearest point algorithms described in Sections~\ref{sec:gluevectoralg},~\ref{sec:onlognalg} and~\ref{sec:onalg} and also some of the specialised algorithms for $A_n$ and $A_n^*$ discussed in Section~\ref{sec:otheralgorithmsAnAn*}.  The algorithms were written in Java and the computer used is a \unit[900]{MHz} Intel Celeron M.

Table~\ref{tab:An/4comp} shows the computation times for the three algorithms from Sections~\ref{sec:gluevectoralg},~\ref{sec:onlognalg} and~\ref{sec:onalg} for the lattice $A_{n/4}$ and $q = \nicefrac{n+1}{4}$.  It is evident that the linear-time algorithm is the fastest.  The glue vector algorithm is significantly slower for large $n$.  By comparison, Table~\ref{tab:An/n+1/4comp} shows the computation times for the algorithms with $A_{n/m}$ for $m = \nicefrac{n+1}{4}$ and $q = 4$.  The glue vector algorithm now performs similarly to the other algorithms.  This behaviour is expected.  As discussed in Section~\ref{sec:gluevectoralg} the glue vector algorithm has linear complexity when $q$ is constant, but quadratic complexity when $q$ increases with $n$.

Tables~\ref{tab:An*comp} and~\ref{tab:Ancomp} show the performance of the linear-time Coxeter lattice algorithm compared to the specialised algorithms for the lattices $A_n^*$ and $A_n$ discussed in Section~\ref{sec:otheralgorithmsAnAn*}.  It is evident that the specialised algorithms are faster.  This behaviour is expected as the specialised algorithms have less computational overhead. 

\begin{table}[htbp] \label{tab:An/4comp}
\centering
\caption{Computation time in seconds for $A_{n/4}$ for $10^5$ trials}
\begin{tabular}{lrrr}
Algorithm & \multicolumn{1}{l}{n=25} & \multicolumn{1}{l}{n=100} & \multicolumn{1}{l}{n=1000} \\ \toprule
$O(n)$ & 6.14 & 18.89 & 165.57   \\ 
$O(n\log{n})$ & 6.83 & 21.51 & 205.36 \\ 
$O(n^2)$ & 13.66 & 161.80 & $>10^4$  \\ \bottomrule
\end{tabular}
\label{tab:computation_time_An/4}
\end{table}

\begin{table}[htbp] \label{tab:An/n+1/4comp}
\centering
\caption{Computation time in seconds for $A_{n/m}$ with $m = \frac{n+1}{4}$ for $10^5$ trials}
\begin{tabular}{lrrr}
Algorithm & \multicolumn{1}{l}{n=25} & \multicolumn{1}{l}{n=100} & \multicolumn{1}{l}{n=1000} \\ \toprule
$O(n)$ & 6.67 & 17.78 & 157.66   \\ 
$O(n\log{n})$ & 21.33 & 9.27 & 209.23 \\ 
$O(n^2)$ & 10.71 & 35.24 & 317.14  \\ \bottomrule
\end{tabular}
\label{tab:computation_time_An/(n+1)/4}
\end{table}

\begin{table}[htbp] \label{tab:An*comp}
\centering
\caption{Computation time in seconds for linear-time $A_n^*$ \cite{McKilliam2008b} and $A_{n/1}$ (\Alg{alg:slow}) with $10^5$ trials}
\begin{tabular}{lrrr}
Algorithm & \multicolumn{1}{l}{n=25} & \multicolumn{1}{l}{n=100} & \multicolumn{1}{l}{n=1000} \\ \toprule
$A_{n/1}$ & 6.55 & 19.83 & 185.21   \\ 
$A_n^*$ & 6.00 & 14.54 & 125.56  \\ \bottomrule
\end{tabular}
\label{tab:computation_time_An*}
\end{table}

\begin{table}[htbp] \label{tab:Ancomp}
\centering
\caption{Computation time in seconds for linear-time $A_n$ (\Alg{alg:Anon}) and $A_{n/n+1}$ (\Alg{alg:slow}) with $10^5$ trials}
\begin{tabular}{lrrr}
Algorithm & \multicolumn{1}{l}{n=25} & \multicolumn{1}{l}{n=100} & \multicolumn{1}{l}{n=1000} \\ \toprule
$A_{n/n+1}$ & 7.24 & 19.37 & 161.02   \\ 
$A_n$ & 4.19 & 10.69 & 85.45  \\ \bottomrule
\end{tabular}
\label{tab:computation_time_An}
\end{table}

\section{Conclusion}

In this paper we have described two new nearest point algorithms for the Coxeter lattices.  The first algorithm is a generalisation of the nearest point algorithm for $A_n^*$ described in \cite{McKilliam2008} and requires $O(n\log{n})$ arithmetic operations.  The second algorithm requires $O(n)$ operations in the worst case.  The second algorithm makes use of a partial sorting procedure called a bucket sort~\cite{Cormen2001} and also the linear-time Rivest-Tarjan selection algorithm~\cite{Blum1973, Floyd1975, Rivest1975, KnuthACP2}.  In Section~\ref{sec:otheralgorithmsAnAn*} we showed how the log-linear and linear-time algorithms for the Coxeter lattices reduce to simple nearest point algorithms for $A_n$ and $A_n^*$ that already exist in the literature~\cite{SPLAG, McKilliam2008, McKilliam2008b}.




\small
\bibliography{../bib/bib}

\end{document}

%% file: defs.tex
\newcommand{\reals}{{\mathbb R}}
\newcommand{\ints}{{\mathbb Z}}
\newcommand{\integers}{{\mathbb Z}}

\newcommand{\NP}{\operatorname{NearestPt}}

\newcommand{\vol}{\operatorname{vol}}
\newcommand{\vor}{\operatorname{Vor}}

\newcommand{\sortindicies}{\operatorname{sortindices}}

\newcommand{\quickpartition}{\operatorname{quickpartition}}
\newcommand{\quickpartitiontwo}{\operatorname{quickpartition2}}

\newcommand{\br}[1]{{\left( #1 \right)}}

\newcommand{\cubr}[1]{{\left\{ #1 \right\}}}

\newcommand{\floor}[1]{{\left\lfloor #1 \right\rfloor}}

\newcommand{\round}[1]{{\left\lfloor #1 \right\rceil}}

\newcommand{\fracpart}[1]{\left\{ #1 \right\}}

\newcommand{\refeqn}[1]{\eqref{#1}}

\newcommand {\Lem}[1] {Lemma~\ref{#1}}

\newcommand {\Alg}[1] {Algorithm~\ref{#1}}

\usepackage[square,comma,numbers,sort&compress]{natbib}

\usepackage{units}
		
\usepackage{booktabs}
		
\usepackage{ifpdf}
\ifpdf
  \usepackage[pdftex]{graphicx}
  \usepackage[naturalnames]{hyperref}
\else
	\usepackage{graphicx}
\fi

\usepackage{amsmath,amsfonts,amssymb, amsthm, bm} 

\usepackage[vlined, linesnumbered]{algorithm2e}
	
\usepackage{mathrsfs}

\newtheorem{definition}{Definition}

\newtheorem{lemma}{Lemma}


%% file: coxeter_np_algorithm.bbl
\begin{thebibliography}{10}

\bibitem{Clarkson1999:Anstar}
I.~V.~L. Clarkson,
\newblock ``An algorithm to compute a nearest point in the lattice ${A}_n^*$,''
\newblock in {\em Applied Algebra, Algebraic Algorithms and Error-Correcting
  Codes}, Marc Fossorier, Hideki Imai, Shu Lin, and Alain Poli, Eds., vol. 1719
  of {\em Lecture Notes in Computer Science}, pp. 104--120. Springer, 1999.

\bibitem{SPLAG}
J.~H. Conway and N.~J.~A. Sloane,
\newblock {\em Sphere packings, lattices and groups},
\newblock Springer, 3rd edition, 1998.

\bibitem{AjtaiGeneratingHardInstancesOfLatticeProblems1996}
M.~Ajtai,
\newblock ``Generating hard instances of lattice problems,''
\newblock {\em in {Proc.} 28th {ACM} {Symposium} on {Theory} of {Computing}},
  pp. 99--108, May 1996.

\bibitem{AjtaiPublicKeyCryptoLattice1997}
M.~Ajtai and C.~Dwork,
\newblock ``A public-key cryptosystem with worst-case/average-case
  equivalence,''
\newblock {\em in {Proc.} 29th {ACM} {Symposium} on {Theory} of {Computing}},
  pp. 284--293, May 1997.

\bibitem{Brunel_lattice_decoding_2003}
L.~Brunel and J.~J. Boutros,
\newblock ``Lattice decoding for joint detection in direct-sequence {CDMA}
  systems,''
\newblock {\em IEEE Trans. Inform. Theory}, vol. 49, pp. 1030--1037, 2003.

\bibitem{Ryan2008}
D.~J. Ryan, I.~V.~L. Clarkson, I.~B. Collings, and .~W. {Heath Jr.},
\newblock ``Performance of vector perturbation multiuser {MIMO} systems with
  limited feedback,''
\newblock Accepted for \emph{IEEE Trans. Commun.}, September 2008.

\bibitem{Conway1982FastQuantDec}
J.~H. Conway and N.~J.~A. Sloane,
\newblock ``Fast quantizing and decoding and algorithms for lattice quantizers
  and codes,''
\newblock {\em IEEE Trans. Inform. Theory}, vol. 28, no. 2, pp. 227--232, Mar.
  1982.

\bibitem{feige_inapproximability_2004}
U.~Feige and D.~Micciancio,
\newblock ``The inapproximability of lattice and coding problems with
  preprocessing,''
\newblock {\em Journal of Computer and System Sciences}, vol. 69, no. 1, pp.
  45--67, Aug 2004.

\bibitem{regev_new_2004}
O.~Regev,
\newblock ``New lattice-based cryptographic constructions,''
\newblock {\em J. {ACM}}, vol. 51, no. 6, pp. 899--942, 2004.

\bibitem{Micciancio_lattice_based_post_quantum_crypto}
D.~Micciancio and O.~Regev,
\newblock ``Lattice based cryptography,''
\newblock in {\em Post Quantum Cryptography}, D~.J. Bernstein, J.~Buchmann, and
  E.~Dahmen, Eds. Springer, 2009.

\bibitem{vanEmdeBoas1981}
P.~{van Emde Boas},
\newblock ``Another {NP}-complete partition problem and the complexity of
  computing short vectors in a lattice,''
\newblock Tech. {R}ep., Mathematisch Instituut, Roetersstraat 15, 1018 WB
  Amsterdam, The Netherlands, Apr. 1981.

\bibitem{Ajtai1998}
M.~Ajtai,
\newblock ``The shortest vector problem in {$L^2$} is {NP}-hard for randomized
  reductions,''
\newblock {\em in {Proc.} 30th {ACM} {Symposium} on {Theory} of {Computing}},
  pp. 10--19, May 1998.

\bibitem{arora_hardness_1993}
S.~Arora, L.~Babai, J.~Stern, and Z.~Sweedyk,
\newblock ``The hardness of approximate optimia in lattices, codes, and systems
  of linear equations,''
\newblock in {\em {IEEE} Symposium on Foundations of Computer Science}, 1993,
  pp. 724--733.

\bibitem{micciancio_worst-case_2004}
D.~Micciancio and O.~Regev,
\newblock ``Worst-case to average-case reductions based on gaussian measures,''
\newblock {\em {SIAM} J. on Computing}, vol. 37, pp. 372--381, 2004.

\bibitem{Agrell2002}
E.~Agrell, T.~Eriksson, A.~Vardy, and K.~Zeger,
\newblock ``Closest point search in lattices,''
\newblock {\em IEEE Trans. Inform. Theory}, vol. 48, no. 8, pp. 2201--2214,
  Aug. 2002.

\bibitem{Viterbo_sphere_decoder_1999}
E.~Viterbo and J.~Boutros,
\newblock ``A universal lattice code decoder for fading channels,''
\newblock {\em IEEE Trans. Inform. Theory}, vol. 45, no. 5, pp. 1639--1642, Jul
  1999.

\bibitem{Pohst_sphere_decoder_1981}
M.~Pohst,
\newblock ``On the computation of lattice vectors of minimal length, successive
  minima and reduced bases with applications,''
\newblock {\em SIGSAM Bull.}, vol. 15, no. 1, pp. 37--44, 1981.

\bibitem{Conway1986SoftDecLeechGolay}
J.~H. Conway and N.~J.~A. Sloane,
\newblock ``Soft decoding techniques for codes and lattices, including the
  {Golay} code and the {Leech} lattice,''
\newblock {\em IEEE Trans. Inform. Theory}, vol. 32, no. 1, pp. 41--50, Jan.
  1986.

\bibitem{McKilliam2008}
R.~G. McKilliam, I.~V.~L. Clarkson, and B.~G. Quinn,
\newblock ``An algorithm to compute the nearest point in the lattice
  ${A}_{n}^*$,''
\newblock {\em IEEE Trans. Inform. Theory}, vol. 54, no. 9, pp. 4378--4381,
  Sep. 2008.

\bibitem{McKilliam2008b}
R.~G. McKilliam, I.~V.~L. Clarkson, W.~D. Smith, and B.~G. Quinn,
\newblock ``A linear-time nearest point algorithm for the lattice
  ${A}_{n}^*$,''
\newblock International Symposium on Information Theory and its Applications,
  2008.

\bibitem{Coxeter1951}
H.S.M. Coxeter,
\newblock ``Extreme forms,''
\newblock {\em Canad. J. Math.}, vol. 3, pp. 391--441, 1951.

\bibitem{Martinet2003}
J.~Martinet,
\newblock {\em Perfect lattices in {Euclidean} spaces},
\newblock Springer, 2003.

\bibitem{Calderbank1986E8TrellisCode}
A.R. Calderbank and N.J.A. Sloane,
\newblock ``An eight-dimensional trellis code,''
\newblock {\em Proc. IEEE}, vol. 74, no. 5, pp. 757--759, May 1986.

\bibitem{Wei1987TrellisCodedModulation}
Lee-Fang Wei,
\newblock ``Trellis-coded modulation with multidimensional constellations,''
\newblock {\em IEEE Trans. Inform. Theory}, vol. 33, no. 4, pp. 483--501, Jul
  1987.

\bibitem{ForneyCosetCodes1}
G.~D.~Forney Jr.,
\newblock ``Coset codes {I}: Introduction and geometrical classification,''
\newblock {\em IEEE Trans. Inform. Theory}, vol. 34, no. 5, pp. 1123--1151, Sep
  1988.

\bibitem{ForneyCosetCodes2}
G.~D.~Forney Jr.,
\newblock ``Coset codes {II}: Binary lattices and related codes,''
\newblock {\em IEEE Trans. Inform. Theory}, vol. 34, no. 5, pp. 1152--1187, Sep
  1988.

\bibitem{Conway1982VoronoiRegions}
J.~Conway and N.~Sloane,
\newblock ``Voronoi regions of lattices, second moments of polytopes, and
  quantization,''
\newblock {\em IEEE Trans. Inform. Theory}, vol. 28, no. 2, pp. 211--226, Mar
  1982.

\bibitem{Postol2002LatticeVideoCompression}
M.S. Postol,
\newblock ``Some new lattice quantization algorithms for video compression
  coding,''
\newblock {\em IEEE Trans. Circuits Systems}, vol. 12, no. 1, pp. 53--60, Jan
  2002.

\bibitem{Secord1989}
N.~Secord and R.~de~Buda,
\newblock ``Demodulation of a {Gosset} lattice code having a spectral null at
  {DC},''
\newblock {\em IEEE Trans. Inform. Theory}, vol. 35, no. 2, pp. 472--477, Mar.
  1989.

\bibitem{Clarkson2007}
I.~V.~L. Clarkson,
\newblock ``Approximate maximum-likelihood period estimation from sparse, noisy
  timing data,''
\newblock {\em IEEE Trans. Signal Process.}, vol. 56, no. 5, pp. 1779--1787,
  May 2008.

\bibitem{McKilliam2007}
R.~G. McKilliam and I.~V.~L. Clarkson,
\newblock ``Maximum-likelihood period estimation from sparse, noisy timing
  data,''
\newblock {\em Proc. Internat. Conf. Acoust. Speech Signal Process.}, pp.
  3697--3700, Mar. 2008.

\bibitem{Clarkson1999}
I.~V.~L. Clarkson,
\newblock ``Frequency estimation, phase unwrapping and the nearest lattice
  point problem,''
\newblock {\em Proc. Internat. Conf. Acoust. Speech Signal Process.}, vol. 3,
  pp. 1609--1612, Mar. 1999.

\bibitem{Quinn2007}
B.~G. Quinn,
\newblock ``Estimating the mode of a phase distribution,''
\newblock {\em Asilomar Conference on Signals, Systems and Computers}, pp.
  587--591, Nov 2007.

\bibitem{McKilliam2009LinearTimeBlockPSK}
R.~G. McKilliam, I.~V.~L. Clarkson, D.~J. Ryan, and I.~B. Collings,
\newblock ``Linear-time block noncoherent detection of {PSK},''
\newblock Accepted for \emph{Proc. Internat. Conf. Acoust. Speech Signal
  Process.}, 2008.

\bibitem{Cormen2001}
T.~H. Cormen, C.~E. Leiserson, R.~L. Rivest, and C.~Stein,
\newblock {\em Introduction to Algorithms},
\newblock MIT Press. and McGraw-Hill, 2nd edition, 2001.

\bibitem{Blum1973}
M.~Blum, R.~W. Floyd, V.~R. Pratt, R.~L. Rivest, and R.~E. Tarjan,
\newblock ``Time bounds for selection,''
\newblock {\em J. Comput. Syst. Sci.}, vol. 7, no. 4, pp. 448--461, 1973.

\bibitem{Floyd1975}
R.~W. Floyd and R.~L. Rivest,
\newblock ``The algorithm {SELECT} - for finding the $i$th smallest of $n$
  elements,''
\newblock {\em Commun. ACM}, vol. 18, no. 3, pp. 173, 1975.

\bibitem{Rivest1975}
R.~W. Floyd and R.~L. Rivest,
\newblock ``Expected time bounds for selection,''
\newblock {\em Commun. ACM}, vol. 18, pp. 165--172, Mar 1975.

\bibitem{KnuthACP2}
D.~E. Knuth,
\newblock {\em The Art of Computer Programming}, vol. Volume 2 (Seminumerical
  Algorithms),
\newblock Addison-Wesley, Reading, Ma., 3rd edition, 1997.

\bibitem{Burger1996}
T.~Burger, P.~Gritzmann, and V.~Klee,
\newblock ``Polytope projection and projection polytopes,''
\newblock {\em The American Mathematical Monthly}, vol. 103, no. 9, pp.
  742--755, Nov 1996.

\end{thebibliography}
